# Visualization of topological shear polaritons in gypsum thin films

Pablo Díaz-Núñez[1,2,†,*], Christian Lanza[3,4,†], Ziwei Wang[1,2], Vasyl G. Kravets[1,2], Jiahua Duan[3,4,5,6], José Álvarez-Cuervo[3,4], Aitana Tarazaga Martín-Luengo[3,4], Alexander N. Grigorenko[1,2], Qian Yang[1,2], Alexander Paarmann[7], Joshua Caldwell[8], Pablo Alonso-González[3,4,*], and Artem Mishchenko[1,2,*]

[1]Department of Physics and Astronomy, University of Manchester, Manchester, UK.
[2]National Graphene Institute, University of Manchester, Manchester, UK.
[3]Department of Physics, University of Oviedo; Oviedo, Spain.
[4]Center of Research on Nanomaterials and Nanotechnology CINN (CSIC–Universidad de Oviedo), El Entrego, Spain.
[5]Center for Quantum Physics, Key Laboratory of Advanced Optoelectronic Quantum Architecture and Measurement (MOE), School of Physics, Beijing, China.
[6]Beijing Key Laboratory of Nanophotonics and Ultrafine Optoelectronic System, Beijing, Beijing Institute of Technology, Beijing, China.
[7]Fritz Haber Institute of the Max Planck Society, Berlin, Germany.
[8]Department of Mechanical engineering, Vanderbilt University, Nashville, USA.

*Corresponding authors: pablo.diaznunez@manchester.ac.uk, pabloalonso@uniovi.es, artem.mishchenko@manchester.ac.uk
[†]These authors contributed equally

## Abstract

Low symmetry crystals have recently emerged as a platform for exploring novel light-matter interactions in the form of hyperbolic shear polaritons. These excitations exhibit unique optical properties such as frequency-dispersive optical axes and asymmetric light propagation and energy dissipation, which arise from the presence of non-orthogonal resonances. However, only non-vdW materials have been demonstrated to support hyperbolic shear polaritons, limiting their exotic properties and potential applications. Here we introduce for the first time novel shear phenomena in low symmetry crystal thin films by demonstrating the existence of elliptical and canalized shear phonon polaritons in gypsum, an exfoliable monoclinic sulphate mineral. Our results unveil a topological transition from hyperbolic shear to elliptical shear polaritons, passing through a canalization regime with strong field confinement. Importantly, we observe a significant slowdown of group velocity, reaching values as low as 0.0005c, highlighting the potential of gypsum for "slow light" applications and extreme light-matter interaction control. These findings expand the application scope of low-symmetry crystals with the benefits that an exfoliable material provides, such as stronger field confinement, tunability, and versatility for its incorporation in complex photonic devices that might unlock new optical phenomena at the nanoscale.

# 1. Introduction

Phonon polaritons (PhPs, light-matter hybrid quasiparticles arising from the coupling of infrared photons with lattice vibrations in polar crystals) in thin layers of van der Waals (vdW) materials have attracted enormous attention in recent years. The highly anisotropic crystal nature of these materials has enabled the visualization of PhPs exhibiting exotic optical phenomena at the nanoscale. For instance, materials like hexagonal boron nitride (h-BN),[1–6] alpha-molybdenum trioxide (α-MoO$_3$),[7–10] or alpha-vanadium pentoxide (α-V$_2$O$_5$)[11–13] display hyperbolic PhPs, characterized by high momentum waves, subdiffractional light confinement, and anisotropic or highly directional light propagation. Such hyperbolicity occurs in the so-called Reststrahlen bands (RBs),[14] where the real part of the permittivity tensor components of the material, $\Re\{\varepsilon_{ii}\}$, along different optical axes have opposite signs. For example, in-plane hyperbolicity in α-MoO$_3$ arises when the in-plane permittivity components $\varepsilon_{xx}$, $\varepsilon_{yy}$ satisfy $\Re\{\varepsilon_{xx}\} \cdot \Re\{\varepsilon_{yy}\} < 0$.[7,9] The hyperbolic regimes in these materials lead to a variety of exotic light propagation phenomena, such as negative reflection[15] and refraction,[16,17] light canalization,[18,19] subdiffractional imaging and focusing,[2,20] and strong coupling.[21–25] Recently, studies on polaritons in low-symmetry crystals have led to the discovery of an entirely new type of hyperbolic PhPs, referred to as hyperbolic shear phonon polaritons, which have been characterized experimentally in beta-gallium oxide (β-Ga$_2$O$_3$)[26,27] and cadmium tungstanate (CdWO$_4$),[28] and theoretically investigated in artificial metasurfaces.[29] Unlike high-symmetry crystals with orthogonal axes, such as α-MoO$_3$, these low-symmetry crystals possess monoclinic structure with a non-orthogonal angle between the crystal axes in the monoclinic plane, allowing the existence of non-orthogonal resonances. As a result, the permittivity tensor of these crystals cannot be diagonalized and contains non-zero off-diagonal components.[30] This gives rise to the key features of shear polaritons, that include asymmetric propagation and asymmetric energy dissipation, as well as rotation of the optical axis as a function of the excitation frequency, and hence, dispersion of the propagation direction of the polariton. These exotic polaritonic properties have opened up new avenues for exploring novel optical phenomena and potential applications in nano-optics,[31,32] and they have even been extended to elastodynamic metasurfaces.[33] However, and in contrast to orthogonal crystals, only non-vdW materials have been demonstrated to support hyperbolic shear polaritons limiting their exotic properties and potential applications. Therefore, fundamental questions remain open, such as whether shear phonon polaritons can exist in thin layers or if non-hyperbolic propagation is possible.

In this work we report the first observation of elliptical shear and canalized shear phonon polaritons (ShPhPs) in thin layers of gypsum (CaSO$_4$·2H$_2$O). By combining real-space nano-imaging and nano-FTIR spectroscopy using s-SNOM (scattering-type Scanning Near-Field Optical Microscopy), we directly visualize the transition from hyperbolic shear to elliptical shear polariton propagation, with shear canalization emerging between these two regimes. Importantly, we observe these phenomena in an exfoliable vdW material, which allows strong polaritonic field confinement in thin layers. These findings open new opportunities for integrating low-symmetry crystals into complex heterostructures, enabling the development of advanced nanophotonic devices using shear light-matter excitations.

## 2. Results and Discussion

**Crystal structure and infrared response of Gypsum**

Gypsum (calcium sulphate dihydrate, $CaSO_4 \cdot 2H_2O$) is one of the most abundant sulphate minerals in nature, with numerous industrial applications ranging from construction[34,35] to agriculture.[36] The schematic in Figure 1A shows its monoclinic crystal structure (space group $I2/a$) with different lattice constants $a = 5.67$ Å, $b = 15.15$ Å, $c = 6.28$ Å, and monoclinic angle $\beta \approx 114°$ between $a$ and $c$ axes.[37–40] The crystal structure of gypsum consists of stacked bilayers along the $b$ axis, with $Ca^{2+}$ cations bound to $SO_4^{2-}$ anionic groups exhibiting a twofold-axis symmetry. Between these layers, there is a bilayer of water molecules coordinated to a $Ca^{2+}$ cation, forming two non-equivalent hydrogen bonds with the sulphate oxygen groups, and asymmetric in the crystal structure. This weak hydrogen bonding layer facilitates the cleavage of gypsum on the (010) plane,[41] allowing the exfoliation of flakes along the monoclinic plane (Figure 1B).

The highly anisotropic crystal structure of gypsum is reflected in its optical properties, particularly in the mid-infrared spectral range, where the fundamental vibrations correspond to either the sulphate groups or the water molecules[39,42–47] (Figure 1C and Figure S1). This work focuses on the asymmetric stretching $v_3$ of $SO_4^{2-}$ between 1100 and 1250 cm$^{-1}$, which exhibits very low infrared transmission (Figure 1C), indicating the presence of a RB. Additionally, we clearly see the optical anisotropy of the monoclinic plane exhibiting non-orthogonal oscillators. Particularly, when the light is polarized parallel to the horizontal long edge of the flake in Figure 1C ($\alpha = 0$) we can identify one phonon centred at 1110 cm$^{-1}$. As we rotate the direction of polarization, we observe a second phonon centred at 1138 cm$^{-1}$, which is maximized when light polarization is closely parallel to the short edge of the flake. The angle between the edges is ~114º, matching that of the monoclinic angle $\beta$ for gypsum. Consequently, the long and short edges of the flake in Figure 1B can be identified as $c$ and $a$ axes of the monoclinic plane, respectively. The description of the dielectric permittivity tensor in the monoclinic plane is complex[30,48,49] but, fortunately, the infrared properties of gypsum have been studied extensively in the past[38,39,42,44–47] and Aronson *et al.*[38] reported its complete dielectric permittivity tensor. To unambiguously characterize the polaritonic regimes in a monoclinic crystal the off-diagonal elements in the permittivity tensor are required to be zero at all frequencies for a coordinate system. However, there is no coordinate system in which $\Re\{\varepsilon_{xy}\} = 0$ at all frequencies. To overcome this limitation, we switch to a frequency-dispersive coordinate system [mnz] by rotating the monoclinic plane by the frequency-dependent angle $\gamma(\omega)$:[26,29]

$$\gamma(\omega) = 0.5 * \tan^{-1}\left(\frac{2\Re\{\varepsilon_{xy}(\omega)\}}{\Re\{\varepsilon_{xx}(\omega)\} - \Re\{\varepsilon_{yy}(\omega)\}}\right) \tag{1}$$

The rotated permittivity tensor is included in Figure 1D, exhibiting $\Re\{\varepsilon_{mn}\} = 0$ and $\Im\{\varepsilon_{mn}\} \neq 0$ and, according to the transmission data, two transverse optical (TO) phonons at ~1110 and ~1138 cm$^{-1}$ for $\varepsilon_{nn}$ and $\varepsilon_{mm}$, respectively. Normal to the monoclinic plane there is a TO phonon at ~1124 cm$^{-1}$ for $\varepsilon_{zz}$. Therefore, we can identify a remarkable variety of narrow-frequency polaritonic regimes based on the signs of $\varepsilon_{mm}$, $\varepsilon_{nn}$, and $\varepsilon_{zz}$ as follows: hyperbolic in-plane type I (1110-1122 cm$^{-1}$), hyperbolic in-plane type II (1122-1138 cm$^{-1}$), elliptical (1138-1172 cm$^{-1}$), hyperbolic in-plane type II (1172-1189 cm$^{-1}$), and hyperbolic in-plane type I (1189-1209 cm$^{-1}$). This description is in excellent agreement with our transmittance measurements, as well as with Raman spectra extracted from the same flake (see Figure S1B).

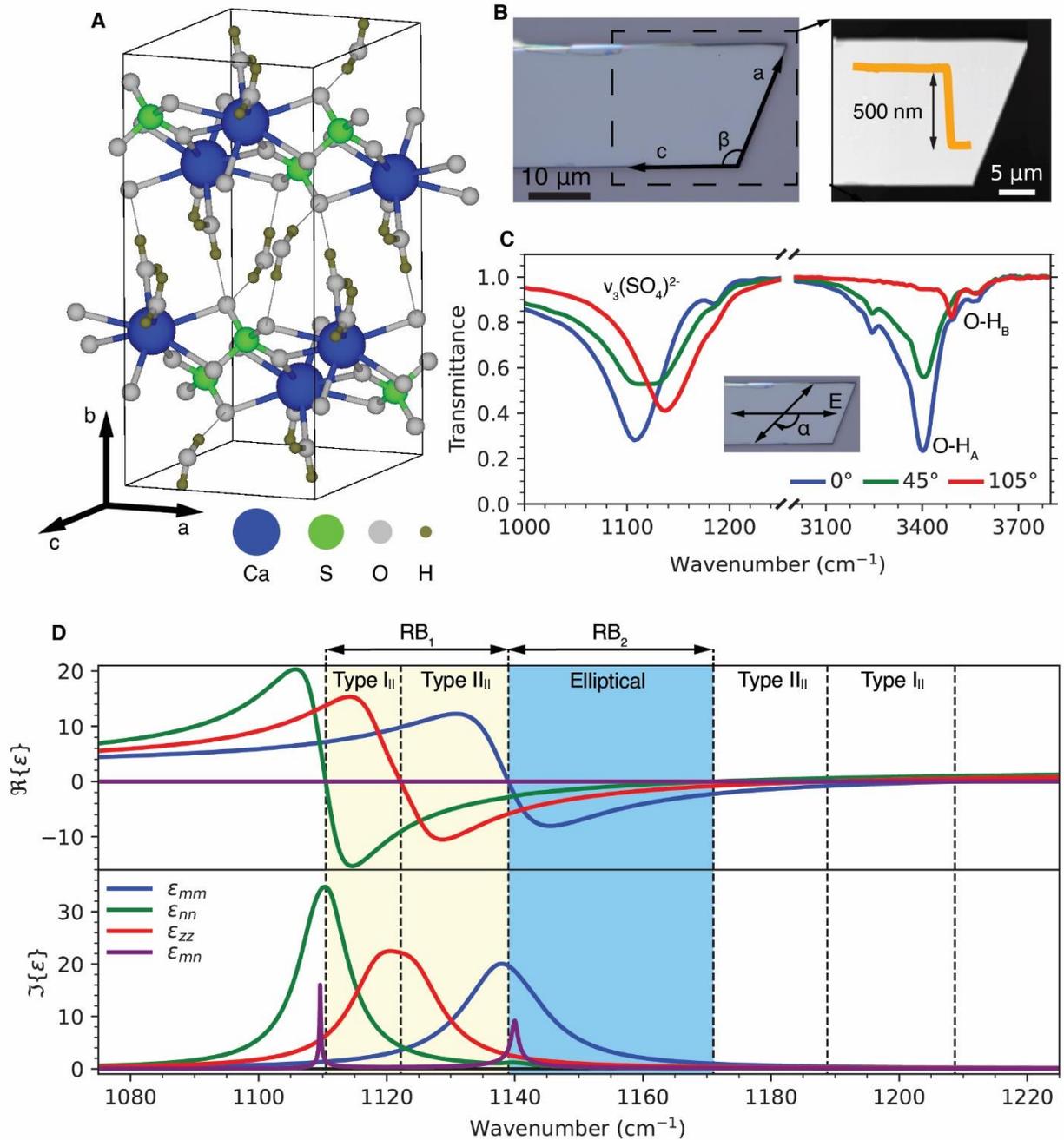

**Figure 1: Crystallographic and optical properties of gypsum.** (**A**) Crystal structure of gypsum. (**B**) Optical image and AFM topography of a 500 nm-thick gypsum flake. (**C**) Infrared transmittance spectrum of the flake in (B) at different polarization angles with respect to the $c$ axis. The inset shows the orientation of the polarization with respect to the gypsum crystal axes: for $\alpha = 0$ the electric field is parallel to the long axis of the flake and is rotating in clockwise direction. (**D**) Dielectric permittivity of gypsum in the frequency-dispersive coordinate system [mnz], exhibiting several RBs.

## Probing phonon polaritons in gypsum thin films

Motivated by these optical findings, we prove the potential excitation of polaritons in gypsum by performing s-SNOM nano-FTIR spectroscopy and real-space imaging on a 75-nm-thick gypsum flake exfoliated on a CaF₂ substrate (Figure 2A). In s-SNOM, the electric field confined at a metallic AFM tip provides enough momenta to launch and probe PhPs in polar crystals. In particular, the launched PhPs propagate until they reflect back at edges or discontinuities, forming a standing wave and generating interference fringes that can be detected by either nano-FTIR spectroscopy or monochromatic nano-imaging.[1,3,7,13,17,50,51]

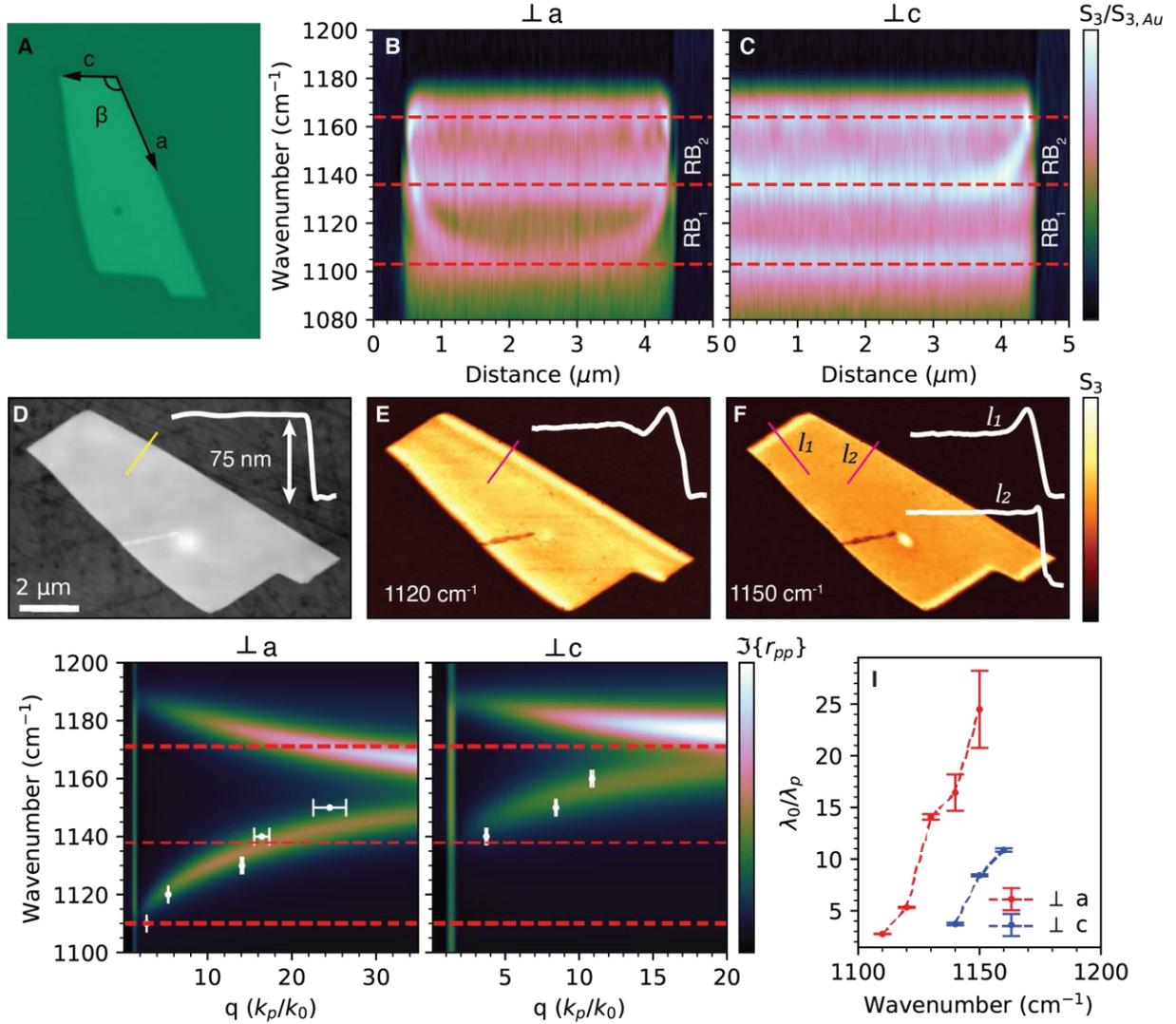

**Figure 2: Near-field optical spectroscopy and imaging of PhPs in gypsum.** (**A**) Optical image of a gypsum flake placed on a CaF₂ substrate. (**B,C**) s-SNOM nano-FTIR line scans perpendicular to *a* and *c* gypsum crystal axes, respectively. (**D**) AFM topography of the gypsum flake with a thickness of 75 nm. (**E,F**) s-SNOM nano-imaging measurements at 1120 and 1150 cm⁻¹, respectively. The insets in (D), (E), and (F) represent the topography and near-field amplitude profiles along the lines in the corresponding panel. (**G,H**) PhP dispersion relation of a 75 nm-thick gypsum flake calculated with TMM perpendicular to *a* and *c* axes, respectively. The experimental values obtained from the fittings of monochromatic s-SNOM images are shown as white dots with their corresponding error bars. (**I**) PhPs confinement factors extracted perpendicular to *a* and *c* axes.

First, we acquired two broad-band nano-FTIR linescans perpendicular to the right and top edges of the flake, corresponding to the *a* and *c* axes of the gypsum crystal, respectively (Figure 2B and Figure 2C). The resulting hyperspectral images show three distinct maxima as a function of distance from the edge at about 1105, 1135, and 1165 cm$^{-1}$, as marked by the dashed lines in Figure 2B and 2C, revealing two RB: RB$_1$ and RB$_2$. The first two maxima at ~1105 and ~1135 cm$^{-1}$ can be identified as the two TO phonons in the monoclinic plane, in good agreement with the infrared transmittance and the dielectric permittivity shown in Figure 1D. The third peak at 1165 cm$^{-1}$ matches the longitudinal optical (LO) phonon along *nn* direction of the rotated frame, separating the elliptical dispersion from that with in-plane type II hyperbolic dispersion (Figure 1D). More importantly, we observe distinct signal maxima with strong spectral dispersion, i.e., resonances that vary with frequency as a function to the distance from the edge. As previously reported, this behaviour unveils the propagation of PhPs,[1,7,13] which we observed in both RB$_1$ (between 1105 and 1135 cm$^{-1}$) and RB$_2$ (between 1135 and 1165 cm$^{-1}$). Interestingly, in RB$_1$, the signal maxima are only present in the scan perpendicular to the *a* axis, suggesting the excitation of in-plane hyperbolic PhPs.[7] In contrast, RB$_2$ exhibits signal maxima along the two in-plane axes, although more pronounced in the scan normal to the *c* axis, which suggests the excitation of in-plane elliptical PhPs.[7] For both RBs, the PhP frequency increases closer to the edge, indicating a positive phase velocity.[1,7,13]

To further analyse these observations, we performed s-SNOM nano-imaging using monochromatic illumination in the spectral range 1100-1190 cm$^{-1}$. The resulting images clearly show the formation of signal maxima (fringes) parallel to the edges of the flakes (Figure 2E,F and Figure S4), confirming the excitation of propagating PhPs in gypsum.[1,3,7,13,17,50,51] In RB$_1$ (Figure 2E), the PhP fringes are only observed propagating perpendicular to the *a* axis, while no detectable fringes are seen perpendicular to the *c* axis. This observation demonstrates the excitation of PhPs with in-plane hyperbolic propagation. In RB$_2$ (Figure 2F), the fringes are visible propagating perpendicular to both edges, although with a clear difference in intensity and width (related to the PhPs wavelength), which confirms the excitation of PhPs with elliptical propagation. From the near-field raster scans we extracted amplitude and phase, $S_n$ and $\varphi_n$, line profiles perpendicular to *a* and *c* axes, and constructed the complex-value near-field signal, $\sigma_n = S_n e^{i\varphi_n}$. We fitted $\sigma_n$ according to:

$$E(x) = A\frac{e^{i2k_p x}}{\sqrt{2x}} + C,\qquad(2)$$

which describes the electric field of a radially propagating damped wave,[52,53] with A, $k_p$ and C as fitting complex-parameters (Figure S5). The extracted values of $k_p$, representing the complex polariton wavevector, are plotted in Figure 2G, and Figure 2H for PhPs propagating perpendicular to the *a* and *c* axes in gypsum, respectively. The dispersion relations of surface polaritons in a 75 nm-thick gypsum flake calculated by the Transfer-Matrix Method (TMM),[54] are also plotted showing a good agreement with the experiment. The polariton wavelength can be calculated as $\lambda_p = 2\pi/\Re(k_p)$. Interestingly, thanks to the thin layer nature of the gypsum flakes, polariton confinements, $\lambda_0/\lambda_p$, as large as 25 and 10 are obtained perpendicular to the *a* and *c* axes, respectively (Figure 2I). In addition, we calculated the group velocity, $v_G = \partial\omega/\partial k$, by fitting the experimental dispersion relation perpendicular to each crystal axis with a power law function of the form $y = ax^b$ and then performing its numerical derivative, and the polariton lifetime as $\tau = L_p/v_G$. We obtained values from $0.005c$ to $0.0005c$ with lifetimes between 2 and 0.6 ps for polaritons propagating perpendicular to the *a* axis in the frequency range between 1110 to 1150 cm$^{-1}$. In the case of polaritons propagating perpendicular to the *c* axis, we obtained group velocities between 0.004$c$ to 0.001$c$ with a lifetime of around 0.8 ps in the frequency range between 1140 to 1160 cm$^{-1}$ (Figure S5). These exceptionally low group velocities suggest the potential of gypsum films to harness "slow light" phenomena, facilitating enhanced light-matter interactions with possible applications in optical signal processing, nanoscale photonic circuits, and infrared sensing technologies.[55–58]

**Topological transition of shear phonon polaritons in gypsum thin films**

To gain more insight on the propagation of PhPs in gypsum, we fabricated an array of Au disks on CaF$_2$ (with diameters of 200, 400, 650, 850 nm and 1 µm), transferred a 150 nm thick gypsum flake on top of them, and performed s-SNOM near-field imaging (Figure 3). This experiment allows us to directly visualize the directional propagation of PhPs, since the circular contour of the disks allows them to be launched or reflected along all possible directions in the plane (in the case where the tip of the s-SNOM acts as launcher).[20,27,32,59] The main experimental results are summarized in Figure 3C (full measurements are shown in Figure S6), where we plot images of the near-field amplitude, $S_n$, demodulated at the $n$=3 harmonic, as a function of illumination frequency and disk diameter. We observe asymmetric polaritonic patterns emanating from the disks that evolve from a hyperbolic to an elliptic contour with increasing frequency. This demonstrates a transition from hyperbolic to elliptical in-plane propagation of PhPs in gypsum. More importantly, the asymmetry observed in the polaritonic patterns unveils a shear behaviour, which, as highlighted in Figure 1, has its origin in the non-zero off-axis components in the gypsum dielectric permittivity tensor.[26,28] In the hyperbolic regime from 1110 to 1135 cm$^{-1}$ we clearly visualize shear asymmetric hyperbolas, which also display a significant rotation of their in-plane propagation as a function of frequency. This observation indicates a continuous dispersion of the optical axes, which has been previously reported for the bulk materials $\beta$-Ga$_2$O$_3$ and CdWO$_4$[26–28] and, more recently, in trigonal ReS$_2$ and ReSe$_2$ waveguide modes at NIR frequencies[60]. Interestingly, when the excitation frequency matches the TO phonon at ~1138 cm$^{-1}$, $\Re\{\varepsilon_{mm}\}\sim0$, the polaritonic signal consists of two parallel fringes (Figure 3C, between 1135 and 1140 cm$^{-1}$), unveiling a canalization effect in which the PhPs propagate along a single in-plane direction. This phenomenon occurs at the transition from hyperbolic to elliptical PhPs contours with increasing frequency (from left to right panels in Figure 3C), thus revealing a 'topological transition' from an opened to a closed isofrequency contour (IFC). Notably, in contrast to previous reports on canalization,[61–66] canalized PhPs in gypsum show an asymmetric field intensity within their flattened wavefronts, thus unveiling an unprecedented shear nature. Our near field images also reveal polaritonic contours consisting of elongated ellipsoids with asymmetric lobes, demonstrating the excitation of elliptical shear PhPs in gypsum (Figure 3C, 1150 cm$^{-1}$), constituting the first observation of this type of highly asymmetric polaritons in a natural material.

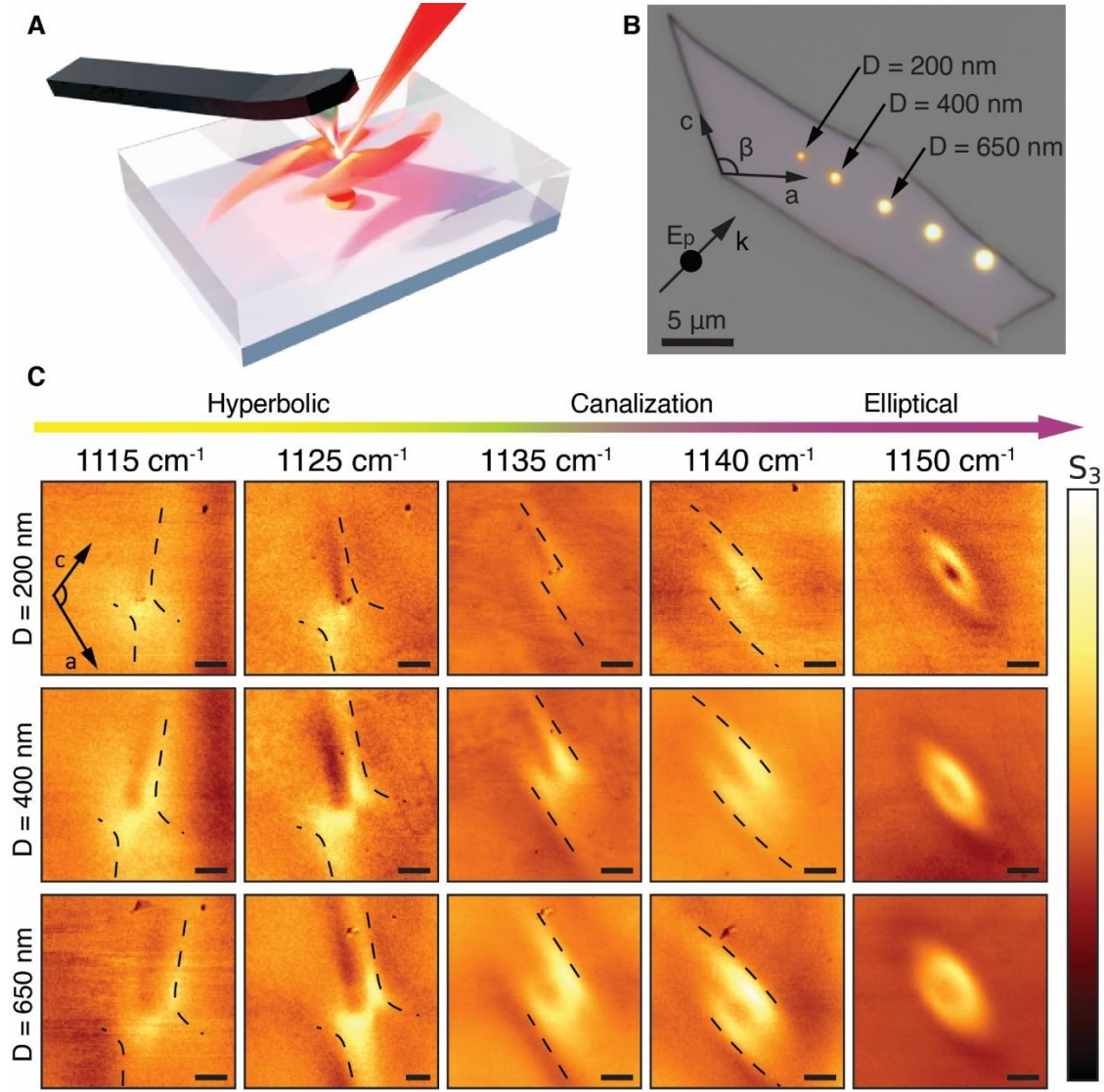

**Figure 3: Near-field imaging of shear phonon polaritons in gypsum: transition from hyperbolic shear to elliptical shear passing through canalized shear propagation of PhPs.** (**A**) Schematic representation of the s-SNOM near-field imaging of PhPs in a gypsum flake placed on top of an Au disk. (**B**) Optical image of the gypsum flake placed on the Au disks array. (**C**) s-SNOM images showing the near-field signal $S_3$ at different incident frequencies (1115, 1125, 1135, 1140, and 1150 cm$^{-1}$, from left to right panels) and for different disk diameters (200, 400, and 650 nm, from top to bottom panels). A clear transition from hyperbolic shear to elliptic shear propagation passing through canalized shear PhPs is observed. The thin dashed black lines are a guide to the eye in the hyperbolic and canalized regime.

To corroborate theoretically the polaritonic response observed in our thin films of gypsum, we conducted an analysis using TMM calculations.[54] First, we verify the polaritonic behaviour of gypsum by analysing the polariton dispersion, $\omega(q)$, in the rotated coordinate system [mnz] along the *mm* (Figure 4A) and *nn* (Figure 4C) axes in the spectral range between 1100 cm$^{-1}$ and 1200 cm$^{-1}$. We clearly observe one polariton branch for each direction starting at different TO phonons. The TO phonon at ~1110 cm$^{-1}$ along the *nn* direction produces the hyperbolic band, RB$_1$, and, at ~1140 cm$^{-1}$, another TO phonon is present along the *mm* direction, leading to the appearance of the elliptical band, RB$_2$, matching our experimental observations. Remarkably, the elliptic RB exhibits a positive phase velocity along both crystal directions, in contrast to previous works,[7,13] where elliptic polaritons showed a negative phase velocity. The dispersion is rather broad in the flat regime, which arises from a large damping value. Our theoretical analysis also confirms the sub-diffractional nature of shear polaritons in gypsum ($q > 1$) along both axes, providing valuable insights into the unique polaritonic characteristics of thin gypsum layers. We also conducted full-wave numerical simulations of shear-polaritons in a thin layer ($d = 150$ nm) of gypsum at four different frequencies: 1120, 1130, 1140, and 1150 cm$^{-1}$, which are shown in Figure 4C-F. Additionally, the IFCs obtained with the TMM method[54] are shown in Figure 4G-J. In agreement with the dispersion shown in Figure 4A and B, and the experimental results shown in Figure 3, we clearly observe hyperbolic shear PhPs at 1120 and 1130 cm$^{-1}$, which display an asymmetric hyperbolic contour. Interestingly, we also observe a canalized shear propagation at 1140 cm$^{-1}$. Such exotic canalization can be understood by examining the IFC, which displays a flat contour with an asymmetric intensity distribution, indicating that the flux of energy points along the same direction for all the allowed wavevectors, but with the asymmetric loss redistribution characteristic of shear polaritons.[27–29,33] This peculiar propagation corresponds to the transition between $\Re\{\varepsilon_{mm}\} > 0$ and $\Re\{\varepsilon_{mm}\} < 0$, which matches the behavior observed in the experimental result shown in Figure 3D. Also, we demonstrate elliptical PhP propagation at 1150 cm$^{-1}$. Finally, to better visualize the topological transition between shear propagation regimes in gypsum we plot in Figure 4K the analytical IFCs (see Methods) in the ($q_x$,$q_y$) plane for q$_z$=0 as a function of frequency. We observe the transition from hyperbolic shear PhPs (open IFCs in yellow) to canalized shear PhPs (dark green) and to elliptical shear PhPs (closed IFCs in blue), in good agreement with our numerical results and, more importantly, with our experimental images in Figure 3.

Interestingly, when comparing the experimental results with the TMM results, we observe that the shear phenomena and the rotation of the hyperbola axis is larger in the experiments than in the calculations. The intensity of the asymmetric propagation and loss redistribution in shear polaritons have been demonstrated to be dependent on two parameters: the relative orientation angle between the oscillators and the losses.[28,29,33] In gypsum we are in a situation where, even though the monoclinic angle is 114°, the angle between the oscillators is about 95°.[38] This angle is very close to 90° and therefore it would induce a small shear effect.[29] However, we need to account for the large optical losses along certain directions in gypsum and, in addition, to the fact that because of the narrow-frequency RBs we are always working close to a TO phonon. The axial dispersion, calculated by $\gamma(\omega)$, varies about 3° between 1110 and 1135 cm$^{-1}$ (Figure S2) and the rotation of the hyperbolic fringes is larger than 3° in the near-field imaging experiments at the same frequency range (Figure 3 and Figure S6). However, the derivation of $\gamma(\omega)$ considers a lossless scenario[29] and it may be valid only when the losses are small.[38,67]

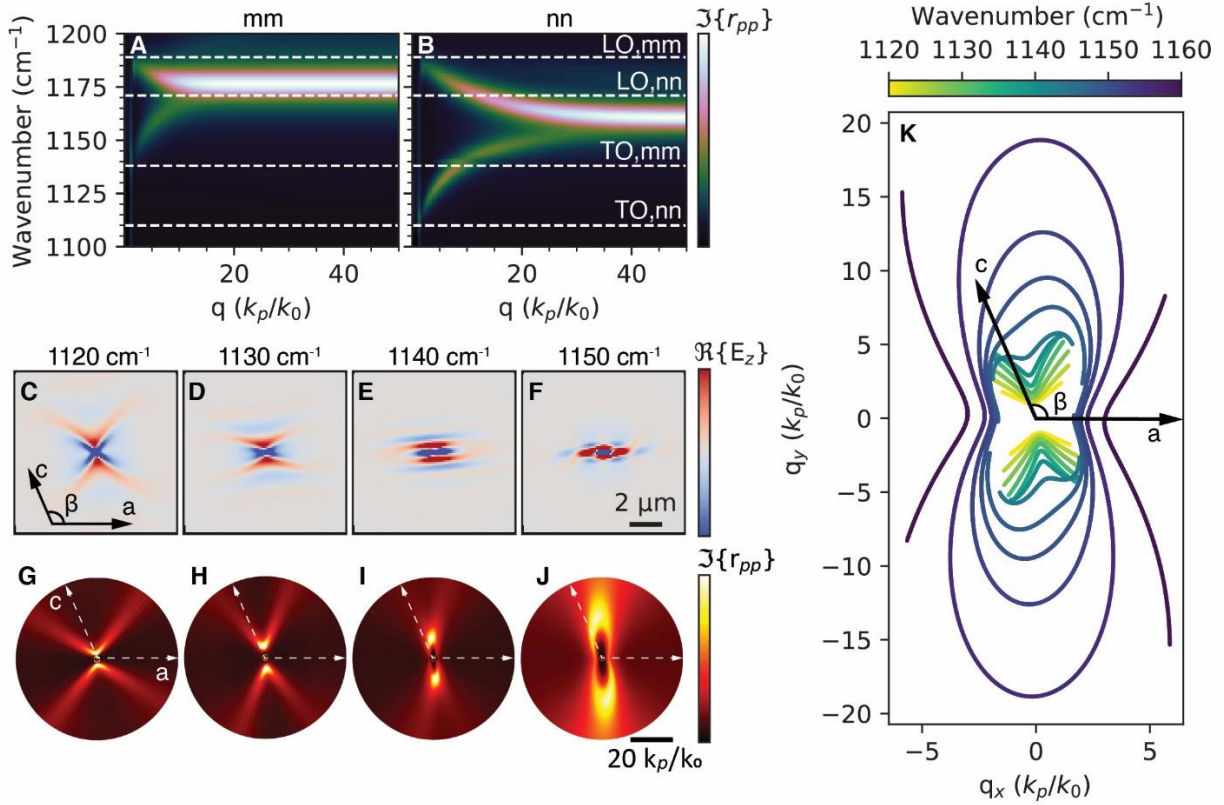

**Figure 4: Theoretical dispersion and topological transition of shear phonon polaritons in thin films of gypsum.** (**A,B**) Dispersion of shear polaritons in 150-nm-thick gypsum along the *mm* (A) and *nn* (B) directions. (**C-E**) Real part of the out-of-plane electric field ($\Re\{E_z\}$) of shear polaritons obtained with numerical simulations at 1120, 1130, 1140 and 1150 $cm^{-1}$. (**G-J**) IFCs obtained with TMM at 1120, 1130, 1140, and 1150 $cm^{-1}$, respectively. (**K**), Analytical IFCs at $q_z = 0$ for PhPs in a 150 nm thick gypsum layer.

## 3. Conclusions

In summary, our work introduces gypsum as a material platform supporting unique polaritonic excitations such as elliptical shear, canalized shear, and hyperbolic shear PhPs. Importantly, these shear optical phenomena are visualized for the first time in thin films of an exfoliable crystal, holding great potential against non-vdW materials in terms of increased field confinement, versatility, and the potential to leverage "slow light" phenomena due to remarkably low group velocities. These distinct polaritonic propagations arise from the very diverse and complex infrared permittivity tensor of gypsum within a narrow frequency range. In addition, the near-field experiments performed in thin layers of gypsum show the expected characteristics from shear polaritons, which include the asymmetric light propagation and loss redistribution, and dispersion of the optical axes. Our work paves the way for its incorporation into complex heterostructures and photonic devices, as the coupling of shear polaritons with plasmons or other phonon polaritons is a promising route to unlock new non-Hermitian optical phenomena at the nanoscale.

# References:

1. Dai, S. *et al.* Tunable Phonon Polaritons in Atomically Thin van der Waals Crystals of Boron Nitride. *Science* **343**, 1125–1129 (2014).
2. Li, P. *et al.* Hyperbolic phonon-polaritons in boron nitride for near-field optical imaging and focusing. *Nat Commun* **6**, 7507 (2015).
3. Giles, A. J. *et al.* Ultralow-loss polaritons in isotopically pure boron nitride. *Nature Mater* **17**, 134–139 (2018).
4. Lee, I.-H. *et al.* Image polaritons in boron nitride for extreme polariton confinement with low losses. *Nat Commun* **11**, 3649 (2020).
5. Menabde, S. G. *et al.* Near-field probing of image phonon-polaritons in hexagonal boron nitride on gold crystals. *Science Advances* **8**, eabn0627 (2022).
6. Caldwell, J. D. *et al.* Sub-diffractional volume-confined polaritons in the natural hyperbolic material hexagonal boron nitride. *Nat Commun* **5**, 5221 (2014).
7. Ma, W. *et al.* In-plane anisotropic and ultra-low-loss polaritons in a natural van der Waals crystal. *Nature* **562**, 557 (2018).
8. Zheng, Z. *et al.* A mid-infrared biaxial hyperbolic van der Waals crystal. *Science Advances* **5**, eaav8690 (2019).
9. Álvarez-Pérez, G. *et al.* Infrared Permittivity of the Biaxial van der Waals Semiconductor α-MoO3 from Near- and Far-Field Correlative Studies. *Advanced Materials* **32**, 1908176 (2020).
10. Wu, Y. *et al.* Chemical switching of low-loss phonon polaritons in α-MoO3 by hydrogen intercalation. *Nat Commun* **11**, 2646 (2020).
11. Clauws, P. & Vennik, J. Lattice Vibrations of V2O5. Determination of TO and LO Frequencies from Infrared Reflection and Transmission. *physica status solidi (b)* **76**, 707–713 (1976).
12. Sucharitakul, S. *et al.* V2O5: A 2D van der Waals Oxide with Strong In-Plane Electrical and Optical Anisotropy. *ACS Appl. Mater. Interfaces* **9**, 23949–23956 (2017).
13. Taboada-Gutiérrez, J. *et al.* Broad spectral tuning of ultra-low-loss polaritons in a van der Waals crystal by intercalation. *Nat. Mater.* **19**, 964–968 (2020).
14. Adachi, S. *Optical Properties of Crystalline and Amorphous Semiconductors*. (Springer US, Boston, MA, 1999). doi:10.1007/978-1-4615-5241-3.
15. Álvarez-Pérez, G. *et al.* Negative reflection of nanoscale-confined polaritons in a low-loss natural medium. *Science Advances* **8**, eabp8486 (2022).
16. Duan, J. *et al.* Planar refraction and lensing of highly confined polaritons in anisotropic media. *Nat Commun* **12**, 4325 (2021).
17. Hu, H. *et al.* Gate-tunable negative refraction of mid-infrared polaritons. *Science* **379**, 558–561 (2023).
18. Li, P. *et al.* Collective near-field coupling and nonlocal phenomena in infrared-phononic metasurfaces for nano-light canalization. *Nat Commun* **11**, 3663 (2020).
19. Duan, J. *et al.* Multiple and spectrally robust photonic magic angles in reconfigurable α-MoO3 trilayers. *Nat. Mater.* **22**, 867–872 (2023).
20. Dai, S. *et al.* Subdiffractional focusing and guiding of polaritonic rays in a natural hyperbolic material. *Nature Communications* **6**, 6963 (2015).
21. Autore, M. *et al.* Boron nitride nanoresonators for phonon-enhanced molecular vibrational spectroscopy at the strong coupling limit. *Light Sci Appl* **7**, 17172–17172 (2018).

22. Bylinkin, A. *et al.* Real-space observation of vibrational strong coupling between propagating phonon polaritons and organic molecules. *Nat. Photonics* **15**, 197–202 (2021).

23. Xu, C., Cai, H. & Wang, D.-W. Vibrational strong coupling between Tamm phonon polaritons and organic molecules. *J. Opt. Soc. Am. B, JOSAB* **38**, 1505–1509 (2021).

24. Dolado, I. *et al.* Remote near-field spectroscopy of vibrational strong coupling between organic molecules and phononic nanoresonators. *Nat Commun* **13**, 6850 (2022).

25. Bylinkin, A. *et al.* Dual-Band Coupling of Phonon and Surface Plasmon Polaritons with Vibrational and Electronic Excitations in Molecules. *Nano Lett.* **23**, 3985–3993 (2023).

26. Passler, N. C. *et al.* Hyperbolic shear polaritons in low-symmetry crystals. *Nature* **602**, 595–600 (2022).

27. Matson, J. *et al.* Controlling the propagation asymmetry of hyperbolic shear polaritons in beta-gallium oxide. *Nat Commun* **14**, 5240 (2023).

28. Hu, G. *et al.* Real-space nanoimaging of hyperbolic shear polaritons in a monoclinic crystal. *Nat. Nanotechnol.* **18**, 64–70 (2023).

29. Renzi, E. M., Galiffi, E., Ni, X. & Alù, A. Hyperbolic Shear Metasurfaces. *Phys. Rev. Lett.* **132**, 263803 (2024).

30. Claus, R. Polariton Dispersion and Crystal Optics in Monoclinic Materials. *physica status solidi (b)* **88**, 683–688 (1978).

31. Krasnok, A. & Alù, A. Low-Symmetry Nanophotonics. *ACS Photonics* **9**, 2–24 (2022).

32. Galiffi, E. *et al.* Extreme light confinement and control in low-symmetry phonon-polaritonic crystals. *Nat Rev Mater* **9**, 9–28 (2024).

33. Yves, S., Galiffi, E., Ni, X., Renzi, E. M. & Alù, A. Twist-Induced Hyperbolic Shear Metasurfaces. *Phys. Rev. X* **14**, 021031 (2024).

34. Lushnikova, N. & Dvorkin, L. 25 - Sustainability of gypsum products as a construction material. in *Sustainability of Construction Materials (Second Edition)* (ed. Khatib, J. M.) 643–681 (Woodhead Publishing, 2016). doi:10.1016/B978-0-08-100370-1.00025-1.

35. Singh, V. K. 13 - Types of gypsum and set regulation of cement. in *The Science and Technology of Cement and Other Hydraulic Binders* (ed. Singh, V. K.) 467–497 (Woodhead Publishing, 2023). doi:10.1016/B978-0-323-95080-0.00013-3.

36. Palacio, S., Azorín, J., Montserrat-Martí, G. & Ferrio, J. P. The crystallization water of gypsum rocks is a relevant water source for plants. *Nat Commun* **5**, 4660 (2014).

37. Pedersen, B. F. & Semmingsen, D. Neutron diffraction refinement of the structure of gypsum, $CaSO_4.2H_2O$. *Acta Cryst B* **38**, 1074–1077 (1982).

38. Aronson, J. R., Emslie, A. G., Miseo, E. V., Smith, E. M. & Strong, P. F. Optical constants of monoclinic anisotropic crystals: gypsum. *Appl. Opt., AO* **22**, 4093–4098 (1983).

39. Anbalagan, G., Mukundakumari, S., Murugesan, K. S. & Gunasekaran, S. Infrared, optical absorption, and EPR spectroscopic studies on natural gypsum. *Vibrational Spectroscopy* **50**, 226–230 (2009).

40. Chen, W. *et al.* Origin of gypsum growth habit difference as revealed by molecular conformations of surface-bound citrate and tartrate. *CrystEngComm* **20**, 3581–3589 (2018).

41. Santos, J. C. C., Negreiros, F. R., Pedroza, L. S., Dalpian, G. M. & Miranda, P. B. Interaction of Water with the Gypsum (010) Surface: Structure and Dynamics from Nonlinear Vibrational Spectroscopy and Ab Initio Molecular Dynamics. *J. Am. Chem. Soc.* **140**, 17141–17152 (2018).

42. Krishnamurthy, N. & Soots, V. Raman Spectrum of Gypsum. *Can. J. Phys.* **49**, 885–896 (1971).

43. Berenblut, B. J., Dawson, P. & Wilkinson, G. R. The Raman spectrum of gypsum. *Spectrochimica Acta Part A: Molecular Spectroscopy* **27**, 1849–1863 (1971).

44. Knittle, E., Phillips, W. & Williams, Q. An infrared and Raman spectroscopic study of gypsum at high pressures. *Phys Chem Min* **28**, 630–640 (2001).

45. Takahashi, H., Maehara, I. & Kaneko, N. Infrared reflection spectra of gypsum. *Spectrochimica Acta Part A: Molecular Spectroscopy* **39**, 449–455 (1983).

46. Prieto-Taboada, N., Gómez-Laserna, O., Martínez-Arkarazo, I., Olazabal, M. Á. & Madariaga, J. M. Raman Spectra of the Different Phases in the CaSO4–H2O System. *Anal. Chem.* **86**, 10131–10137 (2014).

47. Cole, W. F. & Lancucki, C. J. Hydrogen Bonding in Gypsum. *Nature Physical Science* **242**, 104–105 (1973).

48. Mayerhöfer, T. G., Ivanovski, V. & Popp, J. Dispersion analysis of non-normal reflection spectra from monoclinic crystals. *Vibrational Spectroscopy* **63**, 396–403 (2012).

49. Koch, E. E., Otto, A. & Kliewwer, K. L. Reflection spectroscopy on monoclinic crystals. *Chemical Physics* **3**, 362–369 (1974).

50. Chen, J. *et al.* Optical nano-imaging of gate-tunable graphene plasmons. *Nature* **487**, 77–81 (2012).

51. Fei, Z. *et al.* Gate-tuning of graphene plasmons revealed by infrared nano-imaging. *Nature* **487**, 82–85 (2012).

52. Chen, S. *et al.* Real-space observation of ultraconfined in-plane anisotropic acoustic terahertz plasmon polaritons. *Nat. Mater.* **22**, 860–866 (2023).

53. Chen, S. *et al.* Real-space nanoimaging of THz polaritons in the topological insulator Bi2Se3. *Nat Commun* **13**, 1374 (2022).

54. Passler, N. C. & Paarmann, A. Generalized $4 \times 4$ matrix formalism for light propagation in anisotropic stratified media: study of surface phonon polaritons in polar dielectric heterostructures. *J. Opt. Soc. Am. B, JOSAB* **34**, 2128–2139 (2017).

55. Tsakmakidis, K. L., Boardman, A. D. & Hess, O. 'Trapped rainbow' storage of light in metamaterials. *Nature* **450**, 397–401 (2007).

56. Tsakmakidis, K. L., Hess, O., Boyd, R. W. & Zhang, X. Ultraslow waves on the nanoscale. *Science* **358**, eaan5196 (2017).

57. Klein, M. *et al.* Slow light in a 2D semiconductor plasmonic structure. *Nat Commun* **13**, 6216 (2022).

58. Kumar, A. *et al.* Slow light topological photonics with counter-propagating waves and its active control on a chip. *Nat Commun* **15**, 926 (2024).

59. Duan, J., Li, Y., Zhou, Y., Cheng, Y. & Chen, J. Near-field optics on flatland: from noble metals to van der Waals materials. *Advances in Physics: X* **4**, 1593051 (2019).

60. Ermolaev, G. A. *et al.* Wandering principal optical axes in van der Waals triclinic materials. *Nat Commun* **15**, 1552 (2024).

61. Zheng, Z. *et al.* Phonon Polaritons in Twisted Double-Layers of Hyperbolic van der Waals Crystals. *Nano Lett.* **20**, 5301–5308 (2020).

62. Hu, G. *et al.* Topological polaritons and photonic magic angles in twisted α-MoO3 bilayers. *Nature* **582**, 209–213 (2020).

63. Zhou, C.-L., Wu, X.-H., Zhang, Y., Yi, H.-L. & Antezza, M. Polariton topological transition effects on radiative heat transfer. *Phys. Rev. B* **103**, 155404 (2021).

64. Li, S., Zhou, J. & Du, W. Configurable topological phonon polaritons in twisted hBN metasurfaces. *Appl. Opt., AO* **60**, 5735–5741 (2021).

65. Duan, J. *et al.* Enabling propagation of anisotropic polaritons along forbidden directions via a topological transition. *Science Advances* **7**, eabf2690 (2021).

66. Nörenberg, T. *et al.* Germanium Monosulfide as a Natural Platform for Highly Anisotropic THz Polaritons. *ACS Nano* **16**, 20174–20185 (2022).

67. Long, L. L., Querry, M. R., Bell, R. J. & Alexander, R. W. Optical properties of calcite and gypsum in crystalline and powdered form in the infrared and far-infrared. *Infrared Physics* **34**, 191–201 (1993).

68. Álvarez-Pérez, G., Voronin, K. V., Volkov, V. S., Alonso-González, P. & Nikitin, A. Y. Analytical approximations for the dispersion of electromagnetic modes in slabs of biaxial crystals. *Phys. Rev. B* **100**, 235408 (2019).

## Acknowledgements

A.M. and P.D.-N. acknowledge support from the European Research Council (ERC) under the European Union's Horizon 2020 research and innovation program (Grant Agreement No. 865590, Programmable Matter). P.A.-G. acknowledges support from the Spanish Ministry of Science and Innovation (grant PID2022-141304NB-I00) and the European Research Council under Consolidator grant No. 101044461, TWISTOPTICS. J. Duan acknowledges the support from the Beijing Natural Science Foundation (Grant No. Z240005), and National Natural Science Foundation of China (Grant No.12474027).

## Author contributions

P.D.-N. conceived the work together with P.A.-G. P.D.-N., fabricated the devices, and performed nano-FTIR and s-SNOM measurements. J.D. performed s-SNOM measurements. C.L. performed the numerical and analytical study with inputs from A.T.M-L, J.Á,-C., and A.P. P.D.-N. and Z.W. performed the Raman measurements. P.D.-N. and V.K. performed the FTIR measurements. P.D.-N and C.L. wrote the manuscript with input from the rest of authors. All authors contributed to the scientific discussions. A.M. and P. A.-G. supervised the project.

# Supplementary Information

## Materials and Methods

### Sample preparation

Gypsum flakes were mechanically exfoliated onto Si substrates using Nitto tape. The selected flakes were picked up using a polydimethylsiloxane/polypropylene carbonate (PDMS/PPC) stamp fabricated on a glass slide, transferred to a calcium fluoride ($CaF_2$) substrate with prepatterned Au antennas, and cleaned with acetone to eliminate the PPC residue. The antennas were fabricated using standard electron beam lithography with two-layer PMMA resist (495k/950k 3% in anisole for bottom/top layer respectively). The metals (3 nm Cr/27 nm Au) were deposited by e-beam evaporation, followed by a lift-off in acetone, and cleaning with isopropanol.

### Fourier-transformed infrared spectroscopy

Fourier-transform infrared (FTIR) transmittance spectra in the range 800 to 4000 cm$^{-1}$ were recorded with a Bruker Vertex 80 spectrometer equipped with a Bruker Hyperion 3000 FTIR microscope and a nitrogen cooled MCT detector for middle IR region. Gypsum flakes exfoliated on a CaF2 substrate were focused through the standard 20x IR (NA=0.4) Schwarzschild objective. A polarizer (A 675-P) was used to control the angle between the crystal axis and incident polarization direction. The spectra were acquired by rotating the polarizer manually with 15° step. All spectra were normalized to that of the CaF2 substrate, collected with 4 cm$^{-1}$ resolution and averaged over 256 scans.

### Infrared nano-imaging and polariton interferometry

For the nano-imaging experiments, we used an AFM-based commercial scattering-type scanning near-field optical microscope (s-SNOM, Neaspec GmbH). A Pt/Ir coated AFM tip (Arrow™ NCPt, nanoWorld) oscillating at its resonance frequency ($\Omega \approx 285$ kHz) with a tapping amplitude of ~60 nm is illuminated with monochromatic p-polarized light using a tunable QCL with a parabolic mirror while it raster scans the sample. The tip backscattered-light is collected by the same mirror and demodulated at higher harmonics of the tip vibration frequency $n\Omega$ ($n \geq 3$) with a pseudo-heterodyne Michelson interferometer, yielding background-free amplitude and phase images of the near-field signal, together with the sample topography. In near-field imaging, the tip operates as an antenna, enhancing the electric fields in the gap between the tip and providing the necessary momenta distribution to launch and visualize polaritons. In polariton interferometry, the polaritons are either tip- or edge-launched. Tip-launched propagate away from the tip and are back-reflected in the edges of the flake. Edge-launched polaritons are launched by the sharp edges of the flake and travel inwards. In either case interference fringes are generated and measured by the s-SNOM.

The complex near-field signal, $\sigma_n(x) = S_n(x)e^{i\varphi_n(x)}$, extracted from lines perpendicular to $a$ and $c$ axes was fitted according to the equation:

$$E(x) = A\frac{e^{i2k_p x}}{\sqrt{2x}} + C,$$

which describes the electric field of a radially propagating damped wave,[52,53] with A, $k_p$ and C as fitting complex-parameters. $k_p$ represents the complex polariton wavevector. The polariton wavelength can be calculated as $\lambda_p = 2\pi/\Re\{k_p\}$ and the polariton propagation length as $L_p = 1/\Im\{k_p\}$.

Finally, to calculate the group velocity, $v_G = \partial\omega/\partial k$, we fitted the experimental dispersion relation with a power law function of the form $y = ax^b$ and performed its numerical derivative.

The lifetime was calculated as $\tau = L_p/v_G$, where we used the experimental values of the propagation length and the group velocity previously calculated.[13]

**Nano-FTIR spectroscopy**

The nano-FTIR line scans were performed in the same s-SNOM with standard Pt/Ir coated AFM tips (Arrow™ NCPt, nanoWorld). The tip is illuminated with a p-polarized mid-infrared broadband DFG laser (covering a frequency range of about 850–1600 cm$^{-1}$ with an average power of less than 1 mW) using a parabolic mirror. Amplitude and phase spectra are obtained by employing an asymmetric-Fourier transform interferometer on the back-scattered light reflected on the tip. The signal is demodulated to higher harmonics of the tapping amplitude, $n\Omega$ ($n \geq 3$), to achieve background-free detection. We did line scans with a length of 5 µm and a spatial resolution of 25 nm. The nano-FTIR spectra were acquired averaging 3 interferograms of 700 µm (yielding a spectral resolution of about 7 cm$^{-1}$) with 2048 pixels and an integration time of 10 ms/pixel. The recorded spectra were normalized to that of an Au patch evaporated in the same substrate.

**Raman spectroscopy**

Polarized angle-resolved Raman spectra were acquired using a Witec alpha300 R confocal microscope using an excitation wavelength of 532 nm in parallel configuration of the incident and scattered light, i.e., with polarizer and analyser parallel to each other and rotated simultaneously, using a 100x objective and a diffraction grating of 600 grooves/mm. For the measurements in Figure 1, the angle-resolved spectra were recorded with a step of 5°. For each angle, 15 spectra were averaged with a laser power of about 2 mW and integration time of 2 seconds per spectra. For the Raman spectra on Figure S3, the angle-resolved spectra were recorded with a step of 10° using an integration time of 500 s and a laser power of 4 mW for the 75 nm thick flake and 5 mW for the 150 nm thick flake. All the spectra were fitted using a series of Lorentzian functions.

**Transfer-Matrix numerical simulations**

The Transfer-Matrix numerical approach[54] was used to predict the polariton dispersion of a thin layer of gypsum (150 nm) embedded between air and CaF$_2$ as well as the IFCs at several frequencies. We computed the imaginary part of the reflection coefficient ($\Im\{r_{pp}\}$), whose poles determine the maxima of the color plots, corresponding to the polariton dispersion. The permittivity of gypsum has been obtained from[38]. The permittivity of air is assumed to be the vacuum permittivity, while the permittivity of CaF$_2$ is assumed to be 1.6 along the spectral range of interest.

**Full-wave numerical simulations**

The full-wave numerical simulations were performed using the software COMSOL Multiphysics, which is based on the finite boundary elements method. The system is made of 2 quasi semi-infinite media fulfilling the role of superstrate (air) and substrate (CaF$_2$) and a thin layer of the monoclinic crystal gypsum with d = 150 nm. A vertical point dipole placed 100 nm over the surface of the gypsum layer is used as the launcher to excite PhPs.

**Analytical approximations to the dispersion of shear polaritons in thin**

The analytical expression for the dispersion of shear-like polaritons in thin films can be approximated as:

$$q_p = \frac{\rho}{k_0 d}\left[atan\left(\frac{\varepsilon_s \rho}{\varepsilon_{zz}}\right) + atan\left(\frac{\varepsilon_s \rho}{\varepsilon_{zz}}\right) + \pi l\right],$$

where $q_p = \frac{k_p}{k_0}$ stands for the normalized in-plane wavevector, $k_0$ and $d$ are the free-space light wavevector and flake thickness, respectively; $\varepsilon_S$ and $\varepsilon_s$ are the superstrate and substrate permittivity, and

$$\rho = \sqrt{-\frac{\varepsilon_{zz}}{\varepsilon_{xx}cos^2\varphi + \varepsilon_{yy}sin^2\varphi + 2\varepsilon_{xy}sin\varphi cos\varphi}},$$

with $\varphi$ the in-plane angle with regards to the xx axis. A detailed proof of the former result will be given elsewhere.

However, in the diagonal coordinate system [mnz], where the absolute value of the off-diagonal terms $|\Im\{\varepsilon_{mn}\}|$ are small compared to the absolute value of the diagonal terms $|\{\varepsilon_{mm}\}|$ and $|\Im\{\varepsilon_{nn}\}|$ and $|\{\varepsilon_{zz}\}|$, the dispersion of polaritons along the *mm* and *nn* directions can be approximated by the high-momentum approximation[68] with

$$\rho = \sqrt{-\frac{\varepsilon_{zz}}{\varepsilon_{mm/nn}}},$$

for the *mm* and *nn* directions.

**Infrared dielectric permittivity tensor of gypsum**

Aronson *et al.*[38] derived the infrared dielectric permittivity tensor from infrared reflection spectroscopy. They applied a series of Lorentz oscillators to fit the reflectance spectra at three different angles to obtain the dielectric tensor of a monoclinic crystal with incident light normal to the monoclinic plane and parallel to b axis. The general form of the dielectric tensor in a monoclinic crystal with non-zero components in the monoclinic plane (plane xy) in a cardinal system, [xyz], is as follows:

$$\overline{\overline{\varepsilon(\omega)}} = \begin{bmatrix} \varepsilon_{xx}(\omega) & \varepsilon_{xy}(\omega) & 0 \\ \varepsilon_{yx}(\omega) & \varepsilon_{yy}(\omega) & 0 \\ 0 & 0 & \varepsilon_{zz}(\omega) \end{bmatrix},$$

where:

$\varepsilon_{xx}(\omega) = \varepsilon_{xx}^\infty + \sum_k \cos^2(\theta_k) \cdot L_k^{mp}(\omega)$,

$\varepsilon_{xy}(\omega) = \varepsilon_{yx}(\omega) = \varepsilon_{xy}^\infty + \sum_k \cos(\theta_k)\sin(\theta_k) \cdot L_k^{mp}(\omega)$,

$\varepsilon_{yy}(\omega) = \varepsilon_{yy}^\infty + \sum_k \sin^2(\theta_k) \cdot L_k^{mp}(\omega)$,

$\varepsilon_{zz}(\omega) = \varepsilon_{zz}^\infty + \sum_k L_k^{||b}(\omega)$.

Where $\varepsilon^\infty$ are the high frequency permittivity values, $\theta$ is the orientation of the oscillator with respect to the a axis, and $L_k^{mp}$ and $L_k^{||b}$ represent the Lorentz oscillators in the monoclinic plane and parallel to b axis, respectively. Each Lorentzian oscillator, $k$, is defined as:

$$L_k = \frac{S_k}{1 + i\gamma_k\left(\omega/\omega_k\right) - \left(\omega/\omega_k\right)^2},$$

where $\omega_k$ is the resonant frequency (TO phonon), $\gamma_k$ is the damping constant, and $S_k$ is the oscillator strength. The Lorentz oscillators are summarized in Table S1 and Table S2.

**Table S1:** Lorentz oscillator parameters for gypsum in the monoclinic plane.[38]

| Oscillator ($k$) | $\omega_k$ (cm$^{-1}$) | $\gamma_k$ | $S_k$ | $\theta_k$ |
|---|---|---|---|---|
| 1 | 3406.3 | 0.01064 | 0.04313 | 85.04 |
| 2 | 1619.8 | 0.00736 | 0.02079 | 77.59 |
| 3 | 1138.2 | 0.01295 | 0.26264 | 4.93 |
| 4 | 1110.2 | 0.00797 | 0.28429 | 99.87 |
| 5 | 667.4 | 0.01027 | 0.12195 | -7.90 |
| 6 | 598.3 | 0.02559 | 0.16739 | 87.30 |
| 7 | 465.2 | 0.15923 | 0.92815 | -19.05 |
| $\varepsilon_{xx}^{\infty} = 2.2819$ | | $\varepsilon_{yy}^{\infty} = 2.4545$ | | $\varepsilon_{xy}^{\infty} = 0.0086$ |

**Table S2:** Lorentz oscillator parameters for gypsum normal to the monoclinic plane (E ∥ b).[38]

| | $\omega_k$ (cm$^{-1}$) | $\gamma_k$ | $S_k$ |
|---|---|---|---|
| 1 | 3525.8 | 0.01768 | 0.05719 |
| 2 | 1684.5 | 0.00955 | 0.00716 |
| 3 | 1124.3 | 0.00985 | 0.13638 |
| 4 | 1118.6 | 0.00945 | 0.13921 |
| 5 | 601.3 | 0.05583 | 0.24471 |
| 6 | 544.9 | 0.02050 | 0.00798 |
| $\varepsilon_{zz}^{\infty} = 2.223$ | | | |

We reproduce the infrared permittivity of gypsum in Figure S2A in the range 1080-1220 cm$^{-1}$. In the monoclinic plane, the dielectric function features two oscillators at ~1110 and ~1138 cm$^{-1}$ for $\varepsilon_{yy}$ and $\varepsilon_{xx}$, respectively with a non-zero off-axis component ($\varepsilon_{xy}$). Normal to the monoclinic plane there are two overlapping oscillators at ~1118 and ~1124 cm$^{-1}$ for $\varepsilon_{zz}$. In the monoclinic plane, the oscillators at 1138 and 1110 cm$^{-1}$ have an angle of ~5° and ~100° with respect to $a$ axis, respectively. The crystal axis assignment in Aronson´s description[38] of the gypsum crystal structure is equivalent to that used in this work. However, other studies feature an opposite axis notation assignment, which does not affect the overall physical properties. For instance, Takahashi *et al.*[45] assign the first oscillator at 1135 cm$^{-1}$ parallel to the $c$ axis and the second oscillator at 1110 cm$^{-1}$ orthogonal to the $c$ axis.

To capture the rotation of the optical axes in the monoclinic plane in Passler *et al.*[26] diagonalized the real part of the permittivity tensor individually at different frequencies by rotating the monoclinic plane using the frequency-dependent angle:

$$\gamma(\omega) = 0.5 * \tan^{-1}\left(\frac{2\Re\{\mathcal{E}_{xy}(\omega)\}}{\Re\{\mathcal{E}_{xx}(\omega)\} - \Re\{\mathcal{E}_{yy}(\omega)\}}\right)$$

The rotation is performed applying a rotation matrix, $R(\theta)$, defined as:

$$R(\theta) = \begin{bmatrix} \cos\theta & \sin\theta & 0 \\ -\sin\theta & \cos\theta & 0 \\ 0 & 0 & 1 \end{bmatrix}$$

The rotated system, [mnz], is calculated as $\bar{\bar{\varepsilon}}_{[mnz]} = R(-\theta)\bar{\bar{\varepsilon}}_{[xyz]}R(\theta)$ with $\theta = \gamma(\omega)$. In the rotated coordinate system, the real part of $\mathcal{E}_{mn}$ is equal to zero, however, the imaginary part cannot be diagonalized at the same time therefore exhibiting shear phenomena. In the rotated frame we can unambiguously identify the type of polariton propagation supported by gypsum using the real part of the diagonalized permittivity as hyperbolic type I (one element negative), hyperbolic type II (two elements negative), or elliptical (all elements negative). Hyperbolic modes are also classified as in- or out-of-plane.

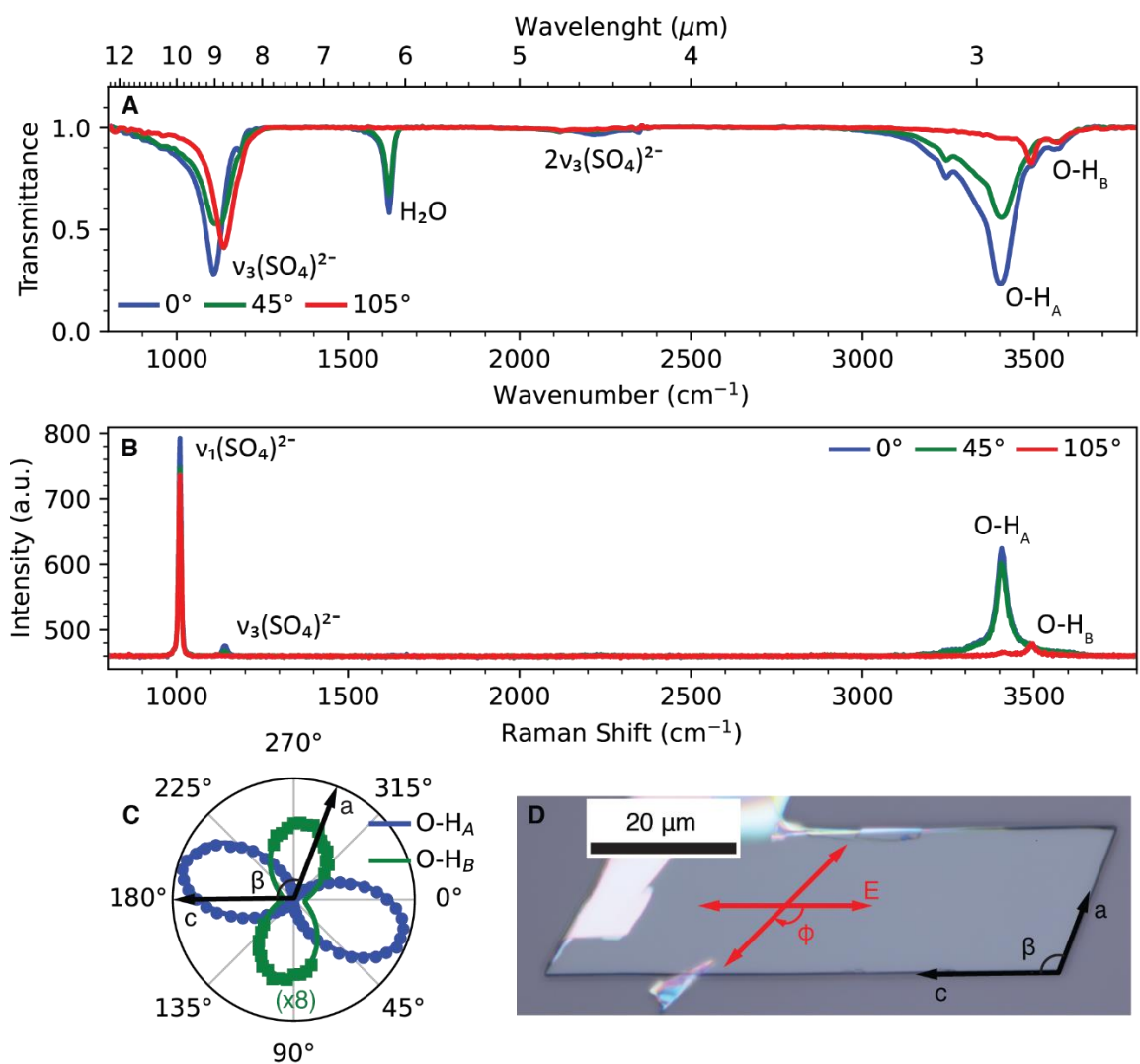

**Figure S1: Infrared response of gypsum in the monoclinic plane**. (**A**) Polarized FTIR transmission and (**B**) Raman spectra in parallel configuration recorded at different angles with respect to the $c$ axis of gypsum. (**C**) Raman polar plot of intensity of the stretching O-H bonds mode of water at ~ 3405 and 3495 cm$^{-1}$. (**D**) Gypsum flake indicating the crystal axis in the monoclinic plane and the polarization. The observed fundamental normal vibrations correspond to the sulphate ionic group, (SO$_4$)$^{2-}$, and the water of crystallization molecules, H$_2$O. Regarding the sulphate vibrations, we first find the Raman-active symmetric stretching ($v_1$) at ~1008 cm$^{-1}$, then the asymmetric stretching ($v_3$) present in both Raman and infrared spectrum between ~1110 and 1140 cm$^{-1}$, and finally the first order overtones of the asymmetric stretching ($2v_3$) between 2100 and 2400 cm$^{-1}$. The water vibrations in gypsum vary from that of the free molecule because it is bonded to the crystal lattice, meaning that the two O-H bonds (O-H$_A$ and O-H$_B$) are not equivalent. In the FTIR spectra we find the infrared active bending mode of water that appears at ~1620 cm$^{-1}$. Note that there is another bending mode at ~1680 cm$^{-1}$ that does not appear in the FTIR spectra in the monoclinic plane because it is an out-of-plane mode. The stretching of the O-H$_A$ and O-H$_B$ modes appear at ~3405 and 3520 cm$^{-1}$.

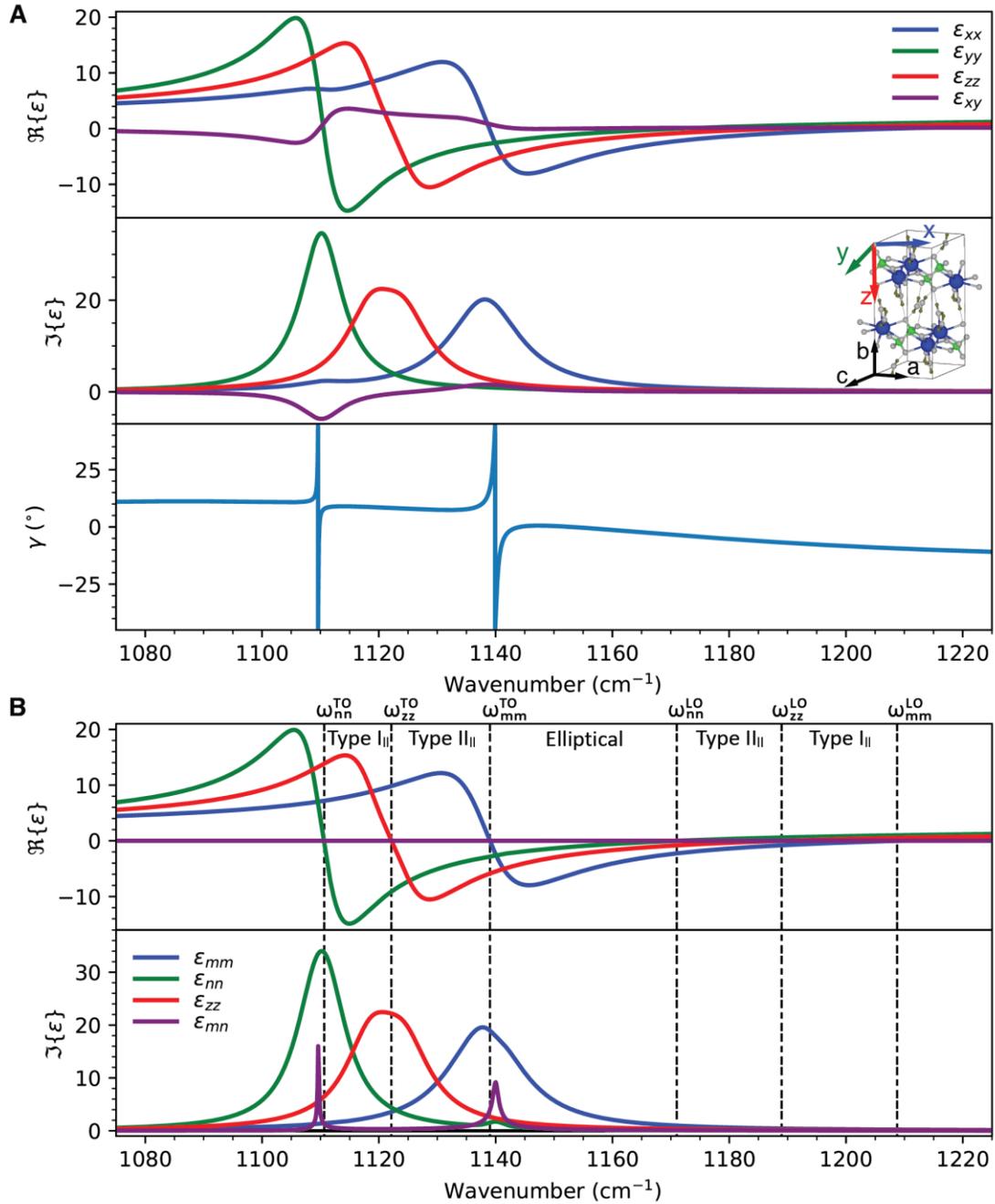

**Figure S2: Infrared permittivity tensor of gypsum in the range between 1075 and 1225 cm⁻¹.**
(**A**) Infrared dielectric permittivity tensor and rotation angle, $\gamma$, of gypsum reproduced from Aronson *et al.*[38] The inset represents the crystal structure with the coordinate system [xyz] in which the permittivity is expressed. (**B**) Dielectric permittivity tensor in the frequency-dispersive coordinate system [mnz] diagonalized by rotating the monoclinic plane by the frequency-dependent rotation angle $\gamma$. The real part of the off-axis components is zero, but the imaginary part is non-zero. Several RBs are identified as a function of the sign of the real part of $\varepsilon_{mm}$, $\varepsilon_{nn}$ and $\varepsilon_{zz}$: type I and type II in-plane hyperbolic polaritons and elliptical polaritons.

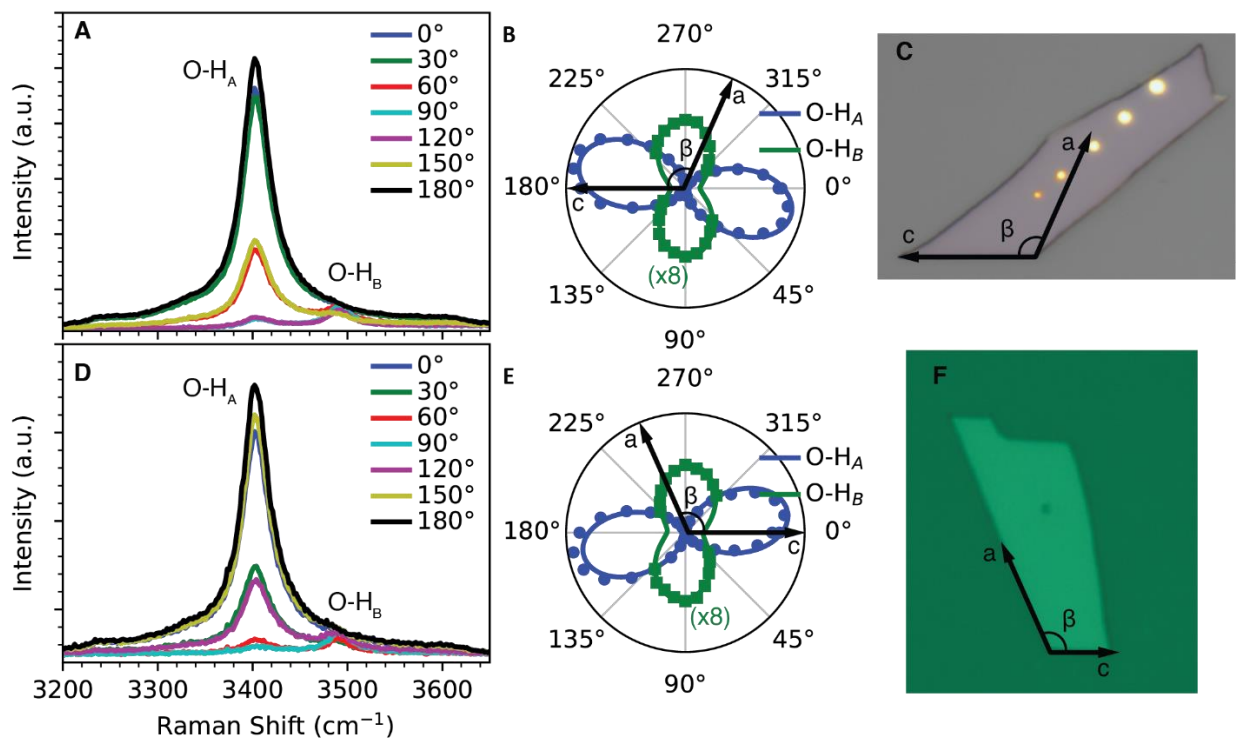

**Figure S3: Gypsum axis assignment based on polarized Raman spectroscopy of the stretching water modes**. (**A,D**) Raman spectra of the stretching mode of water at different angles. (**B,E**) Experimental and fitting of the angular distribution of the O-H$_A$ and O-H$_B$ modes intensity. (**C, F**) Optical images of the gypsum samples indicating the crystal axis. We can identify the crystal axes of gypsum using the water molecules in the crystal structure applying angle-resolved polarized Raman spectroscopy with parallel configuration to probe the Raman-active OH$^-$ stretching vibration. The splitting of the stretching mode of water in gypsum is attributed to two different hydrogen bonds that are present in the crystal structure[47] and well resolved at ~3405 cm$^{-1}$ (O-H$_A$) and ~3480 cm$^{-1}$ (O-H$_B$).[39,41,42,44]

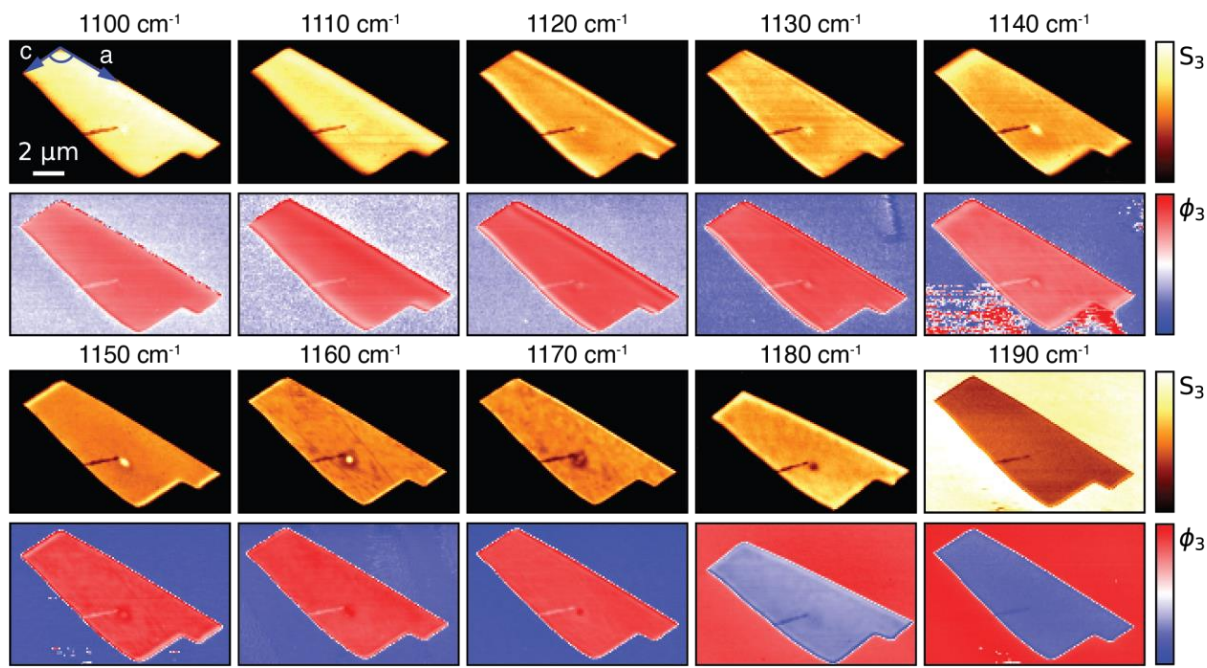

**Figure S4: Polariton interferometry of gypsum in the range 1100-1190 cm⁻¹.** $3^{rd}$ harmonic near-field amplitude ($S_3$) and phase ($\varphi_3$) raster scans of a gypsum flake with a thickness of 75 nm.

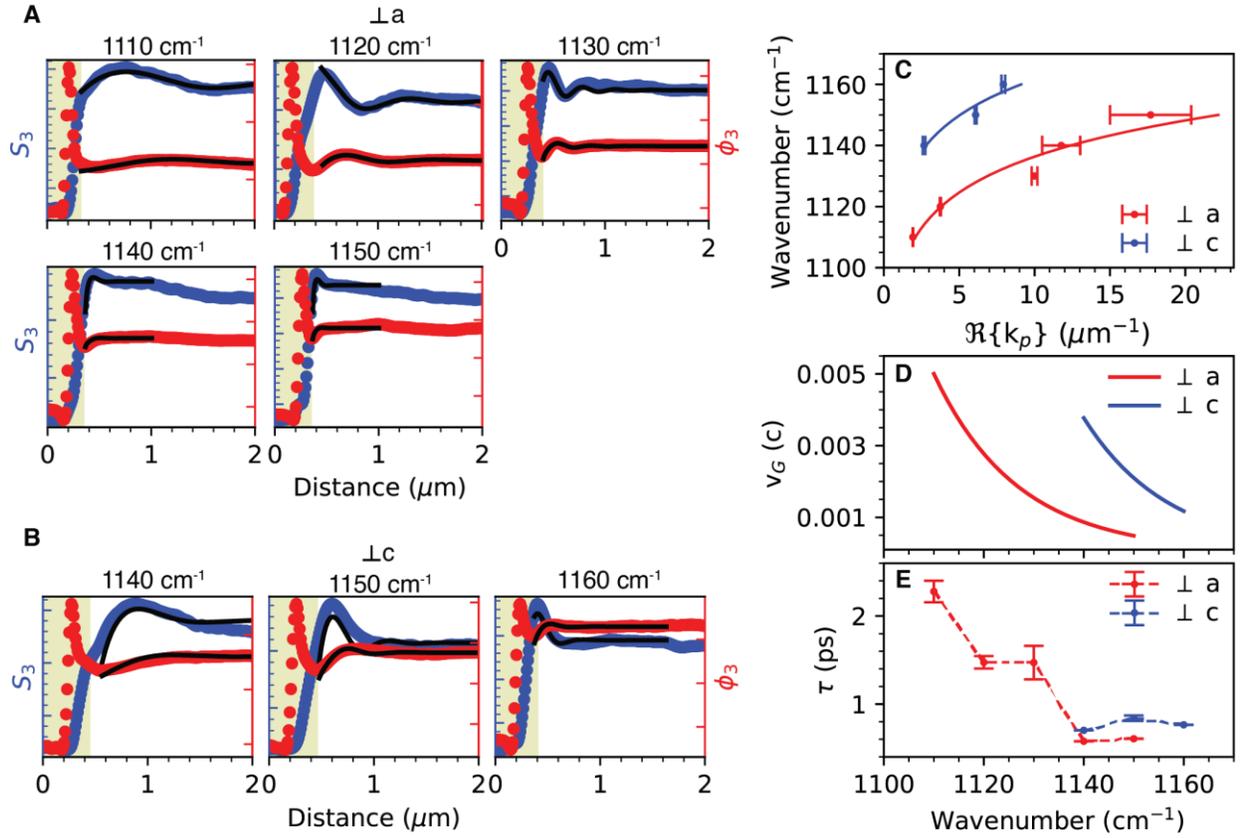

**Figure S5: Analysis of the near-field line profiles perpendicular to gypsum crystal axes.** (**A,B**) 3rd harmonic near-field amplitude (S$_3$) and phase (φ$_3$) line profiles perpendicular to *a* and *c* axes, respectively, at selected frequencies extracted from the near-field images of Figure S4 and fitted according to Eq. 2. The dots represent the experimental measurement and the black lines the best fit. The pale-yellow rectangles in (A) and (B) indicate the substrate position and edge of the flake, which is not taken into account for the fitting. (**C**) Experimental dispersion relation (dots) extracted from the fits in (A) and (B) and fitted with a power law of the form $y = ax^b$ (lines) for polaritons propagating perpendicular to *a* and *c* axes. (**D**) Frequency-dependent group velocity of gypsum polaritons calculated as the numerical derivative of the fitted equation in (C). (**E**) Frequency-dependent polariton lifetime calculated as $\tau = L_p/v_G$.

.

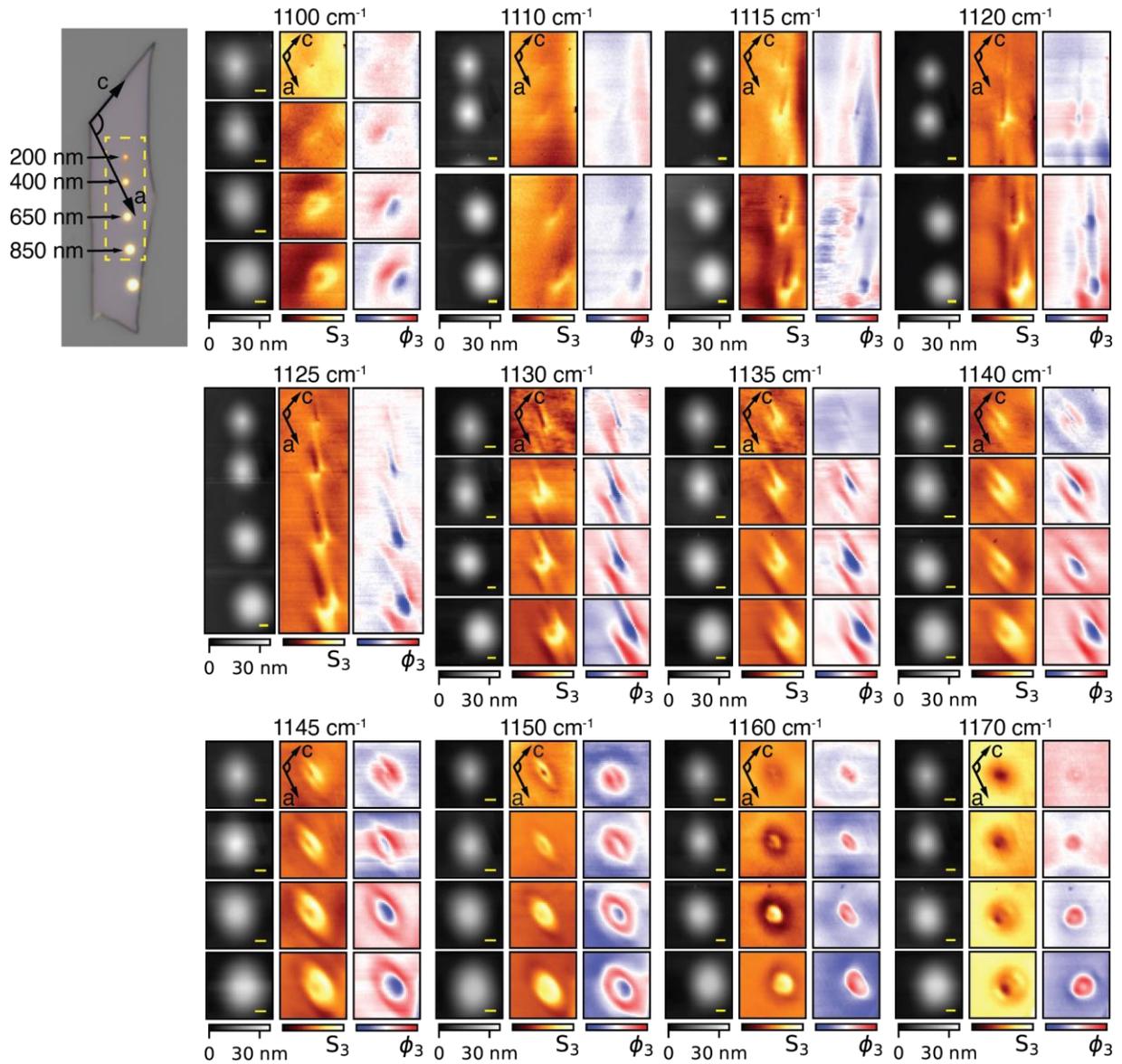

**Figure S6: Real-space observation of shear hyperbolic to shear elliptical transition on gypsum.** Optical image (top left) and s-SNOM measurements on the first four Au disks of increasing diameter in the range 1100-1170 cm$^{-1}$. For each frequency we include the topography, near-field amplitude ($S_3$), and phase ($\varphi_3$). The shear hyperbolic fringes are clearly visualized starting from 1110 cm$^{-1}$ showing the rotation of the fringes as the frequency is increased. The polaritons canalize at 1135-1140 cm$^{-1}$, and close showing shear elliptical polaritons between 1140 to 1150 cm$^{-1}$. The yellow scale bar in the AFM topography maps is 500 nm.

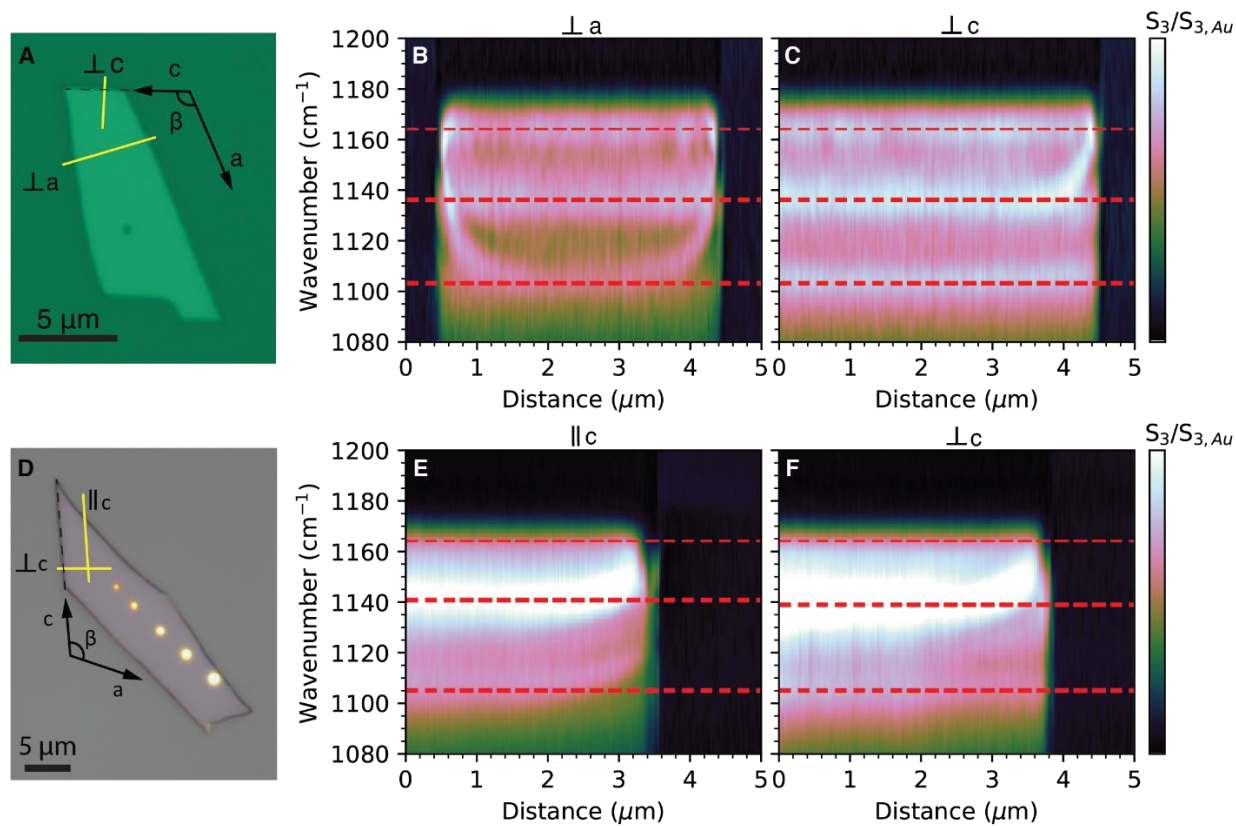

**Figure S7: nano-FTIR spectroscopic line scans of gypsum flakes in the range 1080-1200 cm⁻¹.**
(**A,D**) Optical images of gypsum flakes with a thickness of 75 and 150 nm, respectively. The yellow lines mark the corresponding nano-FTIR spectral line scans. (**B,C**) nano-FTIR line scans perpendicular to the *a* and *c* crystal axes respectively of the flake in (A). (**E,F**) nano-FTIR line scan parallel and perpendicular to the *c* crystal axis respectively of the flake in (D). In the thinnest flake we clearly identify the two RBs mentioned in the main text, reproduced here for comparison. In the thicker flake we can also identify the same regions. Moreover, the TO phonon at 1135 cm⁻¹ is much stronger in relative terms than in the thinner flake and slightly shifted to higher frequencies. In the line scan parallel to the *c* axis, we observe fringes in both RBs, whereas if the scan is performed perpendicular to it, the fringe is observed only in RB₂, in agreement with the measurements in the thinner flake.